\documentclass[12pt,preprintnumbers]{article}
\usepackage{amsmath, amssymb, wasysym, slashed, multirow,hyperref}
\usepackage{pdfpages}
\usepackage{cite}
\usepackage{dsfont}
\usepackage{times}

\newenvironment{sciabstract}{%
\begin{quote} }
{\end{quote}}

\newcounter{lastnote}

\usepackage{fancyhdr}
\fancypagestyle{plain}{%
	\fancyhead[R]{RBI-ThPhys-2020-43,
ACFI-T20-14}

}

\title{On Haag's theorem and renormalization ambiguities}

\author
{Alessio Maiezza,$^{1\ast}$  Juan Carlos Vasquez$^{2\dagger}$\\
\\
\normalsize{$^{1}$Ruder Bo\v skovi\'c Institute, Bijeni\v cka cesta 54, 10000, Zagreb, Croatia,}\\ \\
\normalsize{$^{2}$Amherst Center for Fundamental Interactions, Department of Physics,}\\
\normalsize{University of Massachusetts, Amherst, MA 01003, USA.}\\
\\
\small{ E-mail: amaiezza@irb.hr$^{\ast}$,jvasquezcarm@umass.edu$^{\dagger}$}
}

\date{}

\begin{document}


\baselineskip16pt %24 is the original


\maketitle


\begin{sciabstract}
\section*{Abstract}

We revisit the implications of Haag's theorem in the light  of the renormalization group. There is still some lack of discussion in the literature about the possible impact of the theorem on the standard (as opposite of axiomatic) quantum field theory, and we try to shed light in this direction. Our discussion then deals with the interplay between Haag's theorem and renormalization. While we clarify how perturbative renormalization (for the sub-class of interactions that are renormalizable) marginalizes its impact when the coupling is formally small, we argue that a non-perturbative and non-ambiguous renormalization cannot be built if there is any reference to the interaction picture with free fields. In other words, Haag's theorem should be regarded as a no-go theorem for the existence of a non-ambiguous analytic continuation from  perturbative to  non-perturbative QFT.
\end{sciabstract}

\section{Introduction}

Quantum field theory (QFT) is the merging of quantum mechanics and special relativity through the notion of the quantum field and has been quite successful in describing the fundamental interactions within the standard model -- except gravity. Although QFT has been an incredibly precise tool to calculate observables, particularly in QED, foundational issues are as old as QFT itself. This has brought some authors to try rigorous approaches to formalize the subject, such as
Wightman's program for axiomatic QFT. To date, there is not a completely transparent view of QFT and the related issues on the renormalization.

One problem is that the Stone-von Neumann theorem~\cite{10.2307/1968535,10.2307/1968538}, about the unitary equivalence of the (exponentiated) representations of canonical-commutation-relations (CCR), is not applicable when there are infinite degrees of freedom. Contrary to quantum mechanics, this is indeed the case of the infinite-dimensional space of field theories. Furthermore, when one tries to build the interacting fields from the free ones, canonically quantized, one clashes with the limitations imposed by Haag's theorem~\cite{Haag:1955ev}. The result is a  no-go for the interaction picture with free fields in QFT, upon which perturbation theory is built, and the underlying reason is the vacuum polarization due to interactions. The common belief is that renormalization gives a way out from the no-go (see for example the review in Ref.~\cite{Klaczynski:2016qru}).

The theorem states that a unitary operator, relating the free to the interacting field, does not exist. It prevents, among others, the construction of the Gellmann-Low formula. Needless to say, the latter works very well when computed through the renormalized perturbation theory. In fact, Haag's theorem is often associated with the divergences of perturbation theory and then with renormalization. However, renormalization is a perturbative procedure, and one may worry whether it can actually provide a complete
description of the field evolution and its interactions. In Ref.~\cite{VANHOVE1952145}, Van Hove elaborated in this direction and connected the divergences in QFT with the fact that free and interacting fields do not belong to the same Hilbert space.

In this note, while revisiting Haag's theorem within standard QFT, we provide a fresh look into its consequences within the modern language of the renormalization group and further developed in Refs.~\cite{Maiezza:2019dht,Bersini:2019axn}. Accepting the usual mechanism of absorbing the ultra-violet divergences in the counterterms, we give a self-contained view on how perturbative renormalization marginalizes the impact of Haag's theorem and, more importantly, how the results obtained using perturbative renormalization cannot be analytically continued to finite coupling in a unique way. We then argue that the no-go theorem actually prevents
the construction of a consistent and generic (non-perturbative) formulation of QFT starting from the interaction picture with free fields. In particular, we connect such a no-go to the presence of the renormalon singularities~\cite{tHooft:1977xjm} and elaborate on the link between the resurgence of the renormalization group equation of Refs.~\cite{Maiezza:2019dht,Bersini:2019axn} and the no-go theorem.

\section{Haag's theorem and renormalization}

From the original Haag's work~\cite{Haag:1955ev}, what is currently known as Haag's theorem has been formalized in several versions that either lie in the framework of axiomatic QFT~\cite{hall1957theorem,emch2014algebraic} or rely on an unconventional formalism~\cite{LOPUSZANSKI1962169}.  All these proofs require a substantial amount of effort to get familiar with the formalism used. This is perhaps the main reason why Haag's theorem is completely missing when discussing the Dyson interaction picture in most textbooks on the subject. The pragmatic ignoring of the no-go theorem is also briefly discussed in Ref.~\cite{Hannesdottir:2019opa,Hannesdottir:2019rqq}.

In this note, we prove and clarify the theorem in the language of standard QFT. The proof highlights the \emph{basic} few assumptions needed to prove the theorem and serves as a basis for our main scope, which is dealing with the interplay between Haag's theorem and the language of the renormalization group.

\vspace{0.5cm}

\emph{Haag's theorem.} Let $\phi_0$ and $\phi$ be two real scalar fields, free and interacting,  which act on the Hilbert spaces $\mathcal{H}_0,\mathcal{H}$,  respectively. If at time $t$ they are related with each other through a unitary transformation $U$ and the vacua $|0\rangle, |\Omega\rangle\in \mathcal{H}_0,\mathcal{H} $ are translation-invariant, then the free and interacting vacua must coincide.

\subsection{Revisiting the proof}\label{redo_proof}

\paragraph{Canonical commutation ring.}
Assume the interacting and the free field related by the unitary matrix $U(t,t_0)$ in the following way
\begin{align}
\phi(\textbf{x},t) &  = U(t,t_0)^{-1}  \phi_0(\textbf{x},t) U(t,t_0) \label{phi}\,.
\end{align}
For  models with polynomial interactions only,  the canonical momentum is $\pi= \partial_t\phi$,  and it is not evident how both the field and its canonical momentum transform as in Eq.~\eqref{phi} since
\begin{equation}\label{cond}
\partial_t\phi(x) = U(t,t_0)^{-1} \partial_t \phi_0(x) U(t,t_0) - \left[\phi(x), U(t,t_0)^{-1}\partial_tU(t,t_0)\right].
\end{equation}
The canonical momentum transforms as in Eq.~\eqref{phi} if and only if the last term in Eq.~\eqref{cond} vanishes.
In the Dyson  interaction picture $\phi(x) = \phi_H(x)$ and $\phi_0=\phi_I$ and the matrix $U$  is \emph{unitary} and has the particular  form
\begin{equation}
U(t,t_0)= e^{iH_0(t-t_0)}e^{-iH(t-t_0)}
\end{equation}
where $H$ is the interacting Hamiltonian,  $H_0$ is the free one (both $H$ and $H_0 $ in the Schrodinger picture) and
\begin{equation}
\phi_I(\textbf{x},t) =e^{iH_0(t-t_0)}\phi_H(\textbf{x},t_0) e^{-iH_0(t-t_0)}\,,
\end{equation}
being $\phi_H$ the field in the Heisenberg representation. In this case, it is easy to show that
\begin{align}
U(t,t_0)^{-1}\partial_tU(t,t_0) &= -ie^{i\left(t-t_{0}\right) H}\left(H- H_{0} \right)e^{-i\left(t-t_{0}\right) H}\nonumber \\ &
=-ie^{i\left(t-t_{0}\right) H}H_{int}(t_0) e^{-i\left(t-t_{0}\right) H}=-iH_{int}(t)\,,
\end{align}
thus
\begin{equation}\label{cond2}
 \left[\phi(\textbf{x},t) , U(t,t_0)^{-1}\partial_tU(t,t_0)\right]=0\,,
\end{equation}
and indeed
\begin{equation}\label{pi}
\pi(\textbf{x},t) = U(t,t_0)^{-1}\, \pi_0(\textbf{x},t) \,  U(t,t_0)\,.
\end{equation}
Therefore, the \emph{unitary} Dyson matrix guarantees that both the field and the canonical momentum belong to the same canonical commutation ring. The same conclusion holds if one adds derivative interaction terms since in this case, the second term in Eq.~\eqref{cond} is the required modification of the canonical momentum when more than quadratic time derivatives are present.

\paragraph{Translation invariance.}
Let us denote the free interacting vacua with $|0\rangle$ and $|\Omega \rangle$, upon which two Fock spaces, $\mathcal{H}_0,\mathcal{H}$ respectively, are built. We call $T_0$ and $T$ the translation operators in these two spaces, we assume that under three-dimensional space translations the vacua are translation-invariant and the fields transform as
\begin{align}
& T_0(\textbf{b})^{\dagger}\Phi_0(t,\textbf{x}) T_0(\textbf{b})  = \Phi_0(t,\textbf{x}-\textbf{b})                      \nonumber  \\
& T(\textbf{b})^{\dagger}\Phi(t,\textbf{x}) T(\textbf{b})  = \Phi(t,\textbf{x}-\textbf{b})       \label{translation}                    \,,
\end{align}
where $\Phi=\phi,\pi$.
Plugging  Eq.~\eqref{phi}, where from here on $U$ indicates the unitary Dyson matrix, into Eq.~\eqref{translation}, one gets
\begin{equation} \label{translation1}
U(t,t_0) T(\textbf{b})  =  T_0(\textbf{b})  U(t,t_0) \,,
\end{equation}
and multiplying by $|\Omega\rangle$ to the right
\begin{equation}
U(t,t_0)   |\Omega\rangle=U(t,t_0)T(\textbf{b}) |\Omega\rangle  = T_0(\textbf{b}) U(t,t_0)   |\Omega\rangle    \,,
\end{equation}
where in the last equality we have used the translation-invariance of the interacting vacuum $|\Omega\rangle$.  Since $T_0(\textbf{b})$ acts on states belonging to $\mathcal{H}_0$ and furthermore $U(t,t_0) |\Omega\rangle$ is invariant under translation, then it must be that
\begin{equation}\label{condd1}
U(t,t_0)   |\Omega\rangle =   |0 \rangle \,.
\end{equation}
Similarly, multiplying by $| 0 \rangle$ to the left, one obtains
\begin{equation}
\langle 0 |  U(t,t_0)=\langle 0 | T_0(\textbf{b}) U(t,t_0) = \langle 0 |  U(t,t_0) T (\textbf{b})\,,
\end{equation}
and
\begin{equation}\label{condd2}
 \langle 0 |  U(t,t_0) =  \langle \Omega | \iff  |\Omega\rangle = U^{\dagger }(t,t_0) |0 \rangle\,.
\end{equation}
Finally, from Eqs.~\eqref{condd1} and \eqref{condd2}
\begin{align}
 \langle \Omega | U(t,t_0)  |\Omega\rangle& =  \langle \Omega  |0 \rangle \\
 \langle 0| U^{\dagger}(t,t_0)  |0 \rangle &=  \langle 0  |\Omega \rangle =  \langle \Omega  |0 \rangle^{\dagger} =  \langle \Omega | U^{\dagger}(t,t_0)  |\Omega\rangle\,,
\end{align}
and hence
\begin{equation}\label{same_vacua}
|\Omega \rangle = |0 \rangle\,,
\end{equation}
which concludes the proof~\footnote{In Eqs.~\eqref{condd1} and~\eqref{condd2}, one can also put coefficient $a,b$, with $|a|=|b|=1$, but this does not affect
Eq.~\eqref{same_vacua}}.

It is worth noting that Eq.~\eqref{same_vacua} implies that the two-point correlator for the free and interacting fields are the same:
\begin{align}
\langle \Omega | \mathcal{T}\{ \phi(x)  \phi(y) \} | \Omega \rangle & = \langle 0 | \mathcal{T}\{U^{-1}(x_0,t_0) \phi_I(x)U(x_0,t_0)  U^{-1}(y_0,t_0) \phi_I(y)U(y_0,t_0)  \} | 0 \rangle  =   \nonumber \\
&  =  \langle 0 | \mathcal{T}\{ \phi_I(x)  \phi_I(y) \} | 0 \rangle  \label{same_G}   \,,
\end{align}
having used Eq.~\eqref{phi} inside the time-ordering operator $\mathcal{T}$ and denoting the free field with $\phi_I$ in the conventional notation for interaction picture. Note that the above manipulation is the same one that leads to the Gell-Mann-Low relation, except that in the latter the constraint~\eqref{same_vacua} is not taken into account (see also next Subsection). Eqs.~\eqref{same_vacua} and~\eqref{same_G} make manifest the no-go theorem for interaction picture in QFT, namely the unitary matrix $U$, relating a free and interacting field, does not exist.

\subsection{Dyson operator}

In this subsection, we make explicit the no-go imposed by the above theorem on the construction of the correlation functions within the interaction picture.

One immediate way out might be to postulate that the Dyson operator $U$, relating the free and interacting fields,  is not unitary. However, this is not a possibility since $U$ must be formally unitary. By construction, it obeys the Schr\"odinger-like equation
\begin{equation}\label{ScroedingerEq}
i \partial_t U(t,t_0)= H_I\,  U(t,t_0) \,,
\end{equation}
and the Dyson solution is written as a time-ordered product:
\begin{equation}\label{Dyson_U}
U(t,t_0)=\mathcal{T}\left[\exp\left(-i \int_{t_0}^t H_I\,dt\right)\right]\,.
\end{equation}
The $\mathcal{T}$ operation obscures the unitarity of this matrix and it may be the reason why some authors suggested that a non-unitarity of $U$ should be the key to evade the theorem~\cite{Klaczynski:2013fca}~\footnote{Recall also that the unitarity of $U$ can be rigorously proven in quantum mechanics~\cite{Moretti:2013cma}.}. It is worth stressing that this is not the case, at least formally. The $\mathcal{T}$-product can be disentangled by writing the Magnus solution of Eq.~\eqref{ScroedingerEq}~\cite{doi:10.1002/cpa.3160070404,Iserles1999}
\begin{equation}\label{Magnus}
 U(t,t_0) = \exp\left( -i\int_{t_0}^t H_{int}^I(t_1) dt_1 -\frac{1}{2} \int_{t_0}^t \left[\int_{t_0}^{t_1} H_{int}^I(t_2) dt_2  , H_{int}^I(t_1) \right]dt_1 +...  \right)\,,
\end{equation}
which is manifestly unitary at every order in the expansion \footnote{The drawback is that, at a given order in the expansion, all previous terms have to be computed as well, perhaps making it less useful in practical computations.}. Therefore, the non-unitarity of the Dyson matrix is not an argument to circumvent Haag's theorem. The main point is rather that $U$ does not exist at all. In fact, on the top of (and in agreement with) Haag's theorem, one simply notes that Eq.~\eqref{Magnus} is the solution of Eq.~\eqref{ScroedingerEq} if at least (as a necessary condition; see Ref.~\cite{Blanes_2009} for a review and applications in physics) the operator $U$ belongs to a separable Hilbert space, which is not the case of interacting QFT. However, being expressions~\eqref{Dyson_U} and~\eqref{Magnus} only formal, there is no explicit contradiction with the no-go imposed by Haag's theorem. By using the $S$ matrix
\begin{equation}\label{Smatrix}
S=\lim_{T\rightarrow \infty} U(T,-T)\,,
\end{equation}
to compute the two-point  function (and then calculate the scattering amplitude), one gets
\begin{equation}\label{GellmannLow}
\langle \Omega | \phi(x) \phi(y) | \Omega \rangle =  \frac{\langle 0 |\mathcal{T} \{ S \phi_I(x) \phi_I(y)\} | 0 \rangle}{\langle 0 |\mathcal{T} \{ S\} | 0 \rangle}  \,.
\end{equation}
When  $H=H_0$,  then $S=1$, Eqs.~\eqref{same_G} and~\eqref{GellmannLow} coincide. However, this does not hold in the non-trivial case: the ingredient that makes the difference between Eqs.~\eqref{GellmannLow} and~\eqref{same_G} is to consider $\langle \Omega | 0\rangle\neq 0$ and $|\Omega\rangle\neq |0\rangle$ (see any textbook on QFT, for example Ref.~\cite{Peskin:1995ev}), being the latter in contradiction with Eq.~\eqref{same_vacua}. Nonetheless, Eq.~\eqref{GellmannLow} is in general \emph{explicitly} calculated expanding in power of $H_{int}$, regardless of  the no-go  just discussed.

%\footnote{One needs to invert the limit for $T \rightarrow \infty $ with the power expansion and since this operation is not rigorously justified, the perturbative expansion is in general asymptotic.}.

\vspace{0.5cm}

Our \emph{claim} is: as a consequence of Haag's theorem, an unambiguous calculation of the R.H.S. in Eq.~\eqref{GellmannLow} cannot exist in the interaction picture with free fields, even though renormalized perturbation theory may provide very accurate results in some cases. In what follows, we shall argue in favor of this view and wish to individuate the consequences in QFT, which heavily relies on Eq.~\eqref{GellmannLow}.

\section{Non-perturbative renormalization, renormalization group and resurgence}

In this section, we interpret the above no-go theorem with the non-uniqueness of non-perturbative results, whenever these are derived from the interaction picture with free fields.

It is well-known that the perturbative calculation of the R.H.S. in Eq.~\eqref{GellmannLow} is indeed ill-defined, as the coefficients of powers in the expansion parameter are UV divergent, and certainly, this is not in contradiction with Haag's theorem. \emph{Perturbative renormalization is the solution to this problem}, as long as one formally deals with an infinitesimal coupling (for which perturbation theory is well defined). By using BHP theorem~\cite{bogoliubow1957,Hepp66}, the infinities are reabsorbed in a finite set of counterterms. In this way, perturbative renormalization makes possible an expansion of Eq.\eqref{GellmannLow} through a redefinition of the interaction.

We consider the example of the simple self-interaction of a massless scalar $\phi$ (we focus on UV renormalons, and the same argument holds for any non-asymptotically-free model)
\begin{equation}\label{Lint}
\mathcal{L}_{int} =  \frac{g}{4!} \phi^4  \,.
\end{equation}
At the perturbative level, two counterterms of the form $\delta_Z (\partial\phi)^2$ and $\delta g \phi^4$ are sufficient to reabsorb all infinities appearing at a given order in the perturbative expansion. The choice of $\delta_Z, \delta g$ is not unique, but this issue is kept under control by proper renormalization conditions (and a redefinition of the renormalized coupling $g_R$ within a given scheme). The attempt to go beyond perturbation theory is hampered by the so-called ultraviolet (UV) renormalons. They can be seen as further UV singularities that have to be cured like (and on top of) the original ones appearing in perturbation theory.

\paragraph{Non-perturbative renormalization.}
The important point to be noticed is that Haag's theorem is non-perturbative and thus, for a finite value of the coupling, perturbative renormalization cannot be its solution. Infinities are reabsorbed order-by-order and renormalized perturbation theory provides at best asymptotic series, then rigorously valid for an infinitesimal coupling. To get a complete and self-consistent procedure, one needs to define a procedure valid for finite (or even large) couplings through some further operation on the series, and once at hand,  one would have a non-perturbative-renormalized result. One should regard this hypothetical, rigorous, and complete renormalization at the same level of mathematical robustness as Haag's theorem. In this way, the problem would be evident and sharp, since they would clash with each other. Since Haag's theorem is based on minimal assumptions, as shown in Subsec.~\ref{redo_proof}, the only loophole is that this non-perturbative renormalization, as a well-defined and unambiguous procedure, does not exist.

The above-mentioned operation on the asymptotic series can be achieved through the Borel-Ecalle resummation. However, as shown in Refs.~\cite{Maiezza:2019dht,Bersini:2019axn},  the result obtained using this procedure depends on one arbitrary constant~\footnote{This arbitrary constant makes the result obtained using the Borel-Ecalle resummation ambiguous.} to be fixed by an initial condition for a finite value of the coupling constant.  By the argument of the previous paragraph,  such a condition cannot exist within QFT and therefore we interpret this ambiguity as a direct symptom of Haag's theorem. In what follows, we elaborate on this argument in more detail.

\paragraph{Effective approach via higher-dimensional operators.}
It has been proposed in Ref.~\cite{Parisi:1978iq} that the way to cure the renormalon divergences is to modify the interaction Lagrangian by adding higher dimensional interaction terms, namely
\begin{equation}\label{Lint_mod}
\mathcal{L}_{int} \mapsto \mathcal{L}_{int}+ \sum_{i=3}^{\infty} \delta g_{2 i} \phi^{2 i} \,,
\end{equation}
with $ \delta g_{2 i}= c_{2 i}\,e^{-\frac{1}{2 i \beta_1 g_R}}$ where $\beta_1>0$ is the one-loop beta function. In Eq.~\eqref{Lint_mod}, the \emph{idea} is to use again B.H.P. theorem and the infinite tower of local operators to reabsorb all the UV renormalon singularities. Therefore, one introduces an infinite set of counterterms, i.e. an infinite number of arbitrary constants. On more general ground, the latter conclusion was also given in Refs.~\cite{PhysRev.179.1499,wilson1972}.

In the next section, employing the Borel-Ecalle resummation within the fundamental framework of the renormalization group, the situation is vastly clarified and one can reduce the ambiguity to a single parameter.

\subsection{Renormalization group and resurgence}

Symmetry principles can be used to get non-perturbative information.
In the case under examination, the symmetry is scale transformations. The low energy equation associated with the broken scale symmetry is the so-called Callan-Symanzik equation~\cite{PhysRevD.2.1541,Symanzik:1973pp} (or its ultraviolet approximate version, the RGE). For renormalizable models, this equation provides an alternative way to calculate the coupling constant dependence of physical quantities from the tree level expressions~\cite{PhysRevD.2.1541}. We shall see here that the RGE can be used to obtain a non-linear ordinary differential equation for the nonperturbative corrections of the anomalous dimension. Whereby,  the result depends on one arbitrary constant to be fixed by an initial condition. For renormalizable models, the usual perturbative approach to iteratively solve Green functions using the RGE gives the particular solution to this equation and this is why it appears to be unique. However, once a generalized resummation (generalizing the usual Borel-Laplace) is considered, such as the Borel-Ecalle resummation, one recovers the expected arbitrary constant. Therefore, starting from the interaction picture, the results remain intrinsically ambiguous despite the use of a powerful formalism, and this is in agreement with Haag's theorem.

Recently, the concept of resurgence~\cite{EcalleRes:book} has been proposed as a possible route to connect the perturbative and non-perturbative regimes in QFT~\cite{Dunne:2013ada} in terms of transseries. Interestingly, these can be understood combinatorially (see Ref.~\cite{Mahmoud:2020vww}).
The underlying idea of resurgence is to reduce the number of ambiguities (in Borel transforms) through recursive relations. For the issues discussed above on the renormalon ambiguities, particularly important is the resurgent framework based on ODEs~\cite{Costin1995} used in~\cite{Maiezza:2019dht,Bersini:2019axn}, where it is shown how the infinite ambiguities in Eq.~\eqref{Lint_mod} can be reduced to a single one. We go through the argument in a slightly different way, then we connect with Haag's theorem and the discussion above.

Let us start from the RGE of the two-point Green function
\begin{align}
&[-2 \partial_L+\beta(g)\, \partial_g-2\gamma(g)]G(L,g)=0\,,  \,\,\,\,\,\,\,\,\,  L\equiv \ln(-\frac{p^2}{\mu^2})   \label{RGE} \\
& \langle \Omega | \phi(p) \phi(0) | \Omega \rangle^{-1} = i p^2 G(L,g)  \nonumber     \,.
\end{align}
The crucial point to realize is that if the anomalous dimension $\gamma(g)$ and the beta function $\beta(g)$ were known, the strong dynamics would be also known and therefore QFT would be completely solved (see item (4) page 87 of Ref.~\cite{coleman_1985}) and there would be no need of using Eq.~\eqref{RGE}. In this case the two-point correlator could be  found using the RGE. Since we know from the perturbative renormalization  program that this is not the case, we assume that $\gamma(g),\beta(g)$ are written in terms of a \emph{unknown}, non-perturbative function $\rho(g)$ and define:
\begin{equation}\label{anomalous_dim}
\gamma:=\gamma_{pert}+\rho(g)\,.
\end{equation}
Being $\beta(g)$ related to $\gamma(g)$ it is then a function of $\rho$ and it can be written as
\begin{equation}
\beta=\beta_{pert}+f(\rho(g),g)=\beta_{pert}+q_1 \rho + q_2 \rho g+ q_3 \rho^2 +\mathcal{O}(g^3|\rho^3|g^2 \rho |\rho^2 g)\,,
\end{equation}
where the bars, inside the order of expansion $\mathcal{O}$ denote the logical operator "and/or";
the unknown function $f(\rho(g),g)$ has been expanded for the formally small parameters $g,\rho$, according to the logic of Ref.~\cite{Costin1995}. For
the same reason, we can modify the leading textbook expression for $G\simeq 1 - \gamma_{pert} L$ with a nonperturbative part as
\begin{equation}\label{G}
G(L,g)\simeq1 - \gamma_{pert} L  +k_1 \rho + k_2 \rho g+ k_3 \rho^2 +\mathcal{O}(g^3|\rho^3|g^2 \rho |\rho^2 g)\,.
\end{equation}
The corrections take into account the finite part proportional to $L^0$, which describe the renormalons, and
higher-order perturbative terms in powers of $L$ are irrelevant for our scope~\footnote{In principle, one might also consider power correction in $\rho$ into the coefficient of $L$, but this does not change the conclusions here presented.}. Replacing the expression~\eqref{G} for $G$ in RGE~\eqref{RGE} and expanding for small $g,\rho$, we get the following ODE:
\begin{equation}\label{ODE}
\rho'(g)=\frac{2 \rho }{k_1\beta _1 g^2}-\frac{2 k_2 \rho }{k_1^2\beta _1 g}+\rho ^2 s(g)  +\mathcal{O}(\rho^3) \,.
\end{equation}
The result is that the Borel transform of Eq.~\eqref{ODE}
presents infinite singularities spaced by $2/(k_1\beta_1)$, the coefficient of the leading term $\rho/g^2$ (and the other linear term in $\rho$ characterizes the type of poles). The infinite number of singularities is a consequence of the non-linearity of the ODE, i.e. the term $s(g)\rho^2$, being $s(g)$ a function whose explicit form is irrelevant.

Therefore, matching with the explicit loop estimation of the t'Hooft skeleton diagram, one sets $k_1=1$, and $\rho(g)$ must be identified with a non-perturbative function obtained from the Borel-Ecalle resummation of the renormalons.
In fact, using Eq.~\eqref{anomalous_dim} and Eq.~\eqref{ODE} it is easy to write a similar but non-homogenous equation for the anomalous dimension $\gamma(g)$, the particular solution to this equation $\gamma_{pert}(g)$ is the usual one obtained using the iterative procedure originally proposed in Ref.~\cite{PhysRevD.2.1541}.

The logical point to stress now is that $\gamma_{pert}(g)$ is uniquely determined for (perturbatively) renormalizable models. On the other hand, the homogeneous part
$\rho(g) $ is determined up to an arbitrary constant $C$. Thus, while the renormalons can be formally Borel-Ecalle resummed~\cite{Costin1995} as in Ref.~\cite{Maiezza:2019dht}, the two-point function (and then all Green function related to the former via Swinger-Dyson equations) remains in fact ambiguous. In principle, $C$ should be fixed by an initial condition. The problem is that such condition is not built-in QFT, for if we knew this condition, it would mean the non-perturbative solution is (approximately) known.~\footnote{We should stress again that the approach to RGE, leading to Eq.~\eqref{ODE}, is an approximation and some assumptions are required, such as the non-linearity in $\rho$ are small, in agreement with the theory in Ref.~\cite{Costin1995}}. Note that, even imposing an extra condition, such as UV scale invariance, the constant $C$ remains underdetermined~\cite{Maiezza:2020nbe}. In other words, once the solution is known at a given energy scale, the RGE connects it with a different energy scale, but it cannot help in finding the complete solution.

\vspace{1em}

In summary, although RGE together with the resurgence formalism gives a substantial improvement with respect to the picture outlined in Eq.~\eqref{Lint_mod}, the expression in R.H.S. of Eq.~\eqref{GellmannLow} remains intrinsically ambiguous because of the renormalons. One pays the price to force the $S-$matrix to be meaningful, clashing the no-go imposed by Haag's theorem, with a non-unique, non-universal result. For renormalizable theories, the impact of Haag's theorem should then be understood as the inability to fix the constant $C$  and hence to extend perturbation theory results at the non-perturbative level in a unique way.
Notice also that it provides a rationale on how renormalized perturbation theory can work well despite Haag's theorem: $\gamma_{pert}(g)$ is the \emph{particular} solution to a non-linear differential  equation and it approximates very well the complete solution when $g\rightarrow 0$, as
$\rho(g) $ vanishes faster than $\gamma_{pert}(g)$ for any finite, unknown  value of the constant $C$.

\section{Discussion}

In this note we have elaborated on Haag's theorem, first revisiting the proof in a simplified form based on elementary notions. In this way, we have aimed to make as manifest as possible the inconsistency of the interaction picture upon which standard perturbative QFT is built. Since it is well known that what makes  QFT consistent is renormalization, we have discussed the connection between the theorem with the theory of the renormalization group. The leading theme of this discussion is that, while Haag's theorem is generic and thus non-perturbative, renormalization is often thought of as a perturbative procedure only. It is not surprising that sticking to the latter, the restrictions of  Haag's theorem are quite marginalized (by a finite set of counterterms). However, this cannot be a self-consistent cure, because perturbative renormalization needs to be completed, or in practice resummed. Once this resummation is done, one would have a procedure to make a really consistent QFT for any finite strength of the coupling, but this, we have argued, is prevented indeed by Haag's theorem. \emph{An unambiguous way} to analytically continue renormalization from an infinitesimal coupling (perturbative asymptotic series) to a finite coupling cannot exist. We have identified this with the non-uniqueness of the Borel-Ecalle resummability of the renormalons.

We should also remark that our discussion is focused on usual 4-dimensional QFT, upon which the standard model of particle physics is written. On the other hand, it is worth to comment that within algebraic QFT~\cite{Haag1964} and lower dimensions there are examples, as the Sine-Gordon model, in which the S-matrix power expansion is convergent~\cite{Bahns:2016zqj}. However, to date, there are not known successful descriptions within this approach of realistic models (such as QED).

In summary, we have argued that due to \emph{Haag's theorem}, it is \emph{impossible to uniquely define QFT} starting from the interaction picture with free fields, and the renormalons are the concrete manifestation of this impossibility. These objects are known to do not have a semi-classical limit (i.e. they are deeply at the quantum level because they are a pathology of renormalization), then we have traced back this feature to an ultimate no-go. Note that our discussion has been exemplified with a single coupling model but, since the concept of renormalons can be generalized to multi-coupling~\cite{Maiezza:2018pkk}, we expect that our arguments straightforwardly hold also in this case. Finally, if a non-asymptotically-free model is considered, the suggested implications of Haag's theorem are still valid provided that the subject of the UV-renormalons has to be replaced with the one of IR-renormalons.

%%%%%%%%%%%%%%%%%%%%%%%%%%%%%%%%%%%%%%%%%%%%%%%%%%%%%%%%%%%%%%%%%%%%%%%%%%%%%%%%%%%%%%%%%%%%%%%%%%%%%%%%%%%%%%%%%%%%%%%%%%%%%%%%%%%%%%%%%%%%%%%%%%%%%%%%%%%%%%%%%%%%%%%%%%%%%%%%%%

\section*{Acknowledgement}

We thank Oleg Antipin and Jahmall Bersini for the comments on the manuscript. AM was partially supported by the Croatian Science Foundation project number 4418. JCV was supported in part under the U.S. Department of Energy contract DE-SC0015376.
This is a post-peer-review, pre-copyedit version of an article published in Foundations of Physics. The final authenticated version is available online at: https://link.springer.com/article/10.1007\%2Fs10701-021-00484-3.

%%%%%%%%%%%%%%%%%%%%%%%%%%%%%%%%%%%%%%%%%%%%%%%%%%%%%%%%%%%%%%%%%%%%%%%%%%%%%%%%%%%%%%%%%%%%%%%%%%%%%%%%%%%%%%%%%%%%%%%%%%%%%%%%%%%%%%%%%%%%%%%%%%%%%%%%%%%%%%%%%%%%%%%%%%%%%%%%%%%%

\bibliographystyle{jhep}
\bibliography{biblio}

\end{document}